\documentclass{bmcart}

\usepackage[utf8]{inputenc} 

\usepackage{graphicx}
\usepackage[english]{babel}
\usepackage{amsmath}
\usepackage{mathrsfs}
\usepackage{amsfonts}

\startlocaldefs
\newcommand{\figref}[1]{Fig.~\ref{fig:#1}}

\endlocaldefs

\begin{document}

\begin{frontmatter}

\begin{fmbox}
\dochead{Research}


\title{Quantum Lattice Boltzmann is a quantum walk}

\author[
   addressref={aff1},                   
   corref={aff1},                       
   email={succi@iac.rm.cnr.it}   
]{\fnm{Sauro} \snm{Succi}}
\author[
   addressref={aff3},
   email={francois.fillion@emt.inrs.ca}
]{\fnm{Fran\c{c}ois} \snm{Fillion-Gourdeau}}
\author[
   addressref={aff2},
   email={silviapalpacelli@gmail.com}
]{\fnm{Silvia} \snm{Palpacelli}}


\address[id=aff1]{
  \orgname{IAC-CNR, Istituto per le Applicazioni del Calcolo “Mauro Picone” }, 
  \street{Via dei Taurini, 19},                     %
  \postcode{00185}                                
  \city{Rome},                              
  \cny{Italy}                                    
}
\address[id=aff2]{%
  \orgname{Numidia s.r.l.},
  \street{via G. Peroni, 130},
  \postcode{00131}
  \city{Rome},
  \cny{Italy}
}
\address[id=aff3]{%
  \orgname{Universit\'{e} du Qu\'{e}bec, INRS-\'{E}nergie, Mat\'{e}riaux et T\'{e}l\'{e}communications},
  \street{1650, boulevard Lionel-Boulet},
  \postcode{J3X 1S2}
  \city{Varennes},
  \cny{Canada}
}


\begin{artnotes}
\end{artnotes}

\end{fmbox}


\begin{abstractbox}

\begin{abstract} 
Numerical methods for the 1-D Dirac equation based on operator splitting and on the quantum lattice Boltzmann (QLB) schemes are reviewed. It is shown that these discretizations fall within
the class of quantum walks, i.e. discrete maps for complex fields, whose continuum limit
delivers Dirac-like relativistic quantum wave equations. The correspondence between the quantum walk dynamics and these numerical schemes is given explicitly, allowing a connection between quantum computations, numerical analysis and lattice Boltzmann methods. The QLB method is then extended to the Dirac equation in curved spaces and it is demonstrated that the quantum walk structure is preserved. Finally, it is argued that the existence of this link between the discretized Dirac equation and quantum walks may be employed to simulate relativistic quantum dynamics on quantum computers.
\end{abstract}




\end{abstractbox}
%

\end{frontmatter}



%

\section{Introduction}

A quantum walk (QW) is defined as the quantum analogue of a classical random walk, where the ``quantum walker'' is in a superposition of states instead of being described by a probability distribution. 
One of the earliest realization of this concept was proposed by Feynman as a
discrete version of the massive Dirac equation \cite{FEY}.
In recent times, there has been a surge of interest for this topic, due
to the conceptual and possible practical import of QW's as discrete realizations
of stochastic quantum processes and because they can solve certain problems with an exponential speedup, i.e. using exponentially less operations than classical computations \cite{childs2003exponential}.
Moreover, QW's are amenable to a number of experimental realizations, such as ion traps \cite{PhysRevLett.103.090504,PhysRevLett.104.100503,gerritsma2010quantum}, liquid-state nuclear-magnetic-resonance quantum-information processor \cite{PhysRevA.72.062317},
photonic devices \cite{aspuru2012photonic} and other types of optical devices \cite{QWE}.
As a result, they hold promise of playing an important role in many areas of modern physics
and quantum technology, such as quantum computing, foundational quantum mechanics and biophysics \cite{QW4}.

One of the most interesting features of discrete QW's is their continuum limit, which recovers a broad variety of relativistic quantum wave equations \cite{QW1,QW2,QW3}. As stated earlier, this was first discussed by Feynman and is now known as Feynman's checkerboard \cite{feyn_chess}. This was originally formulated for the free Dirac equation but extensions of these ideas, which include the coupling to external fields, have been investigated \cite{QW1,QW2,QW3}. Pioneering work have been performed in which the Dirac equation is related to cellular automata \cite{PhysRevD.49.6920,meyer1996}.  Lately, the link between QW's and the Dirac equation have been discussed extensively \cite{QW1,QW2,QW3,1751-8121-47-46-465302}.  In these studies, the starting point is a general QW formulation from which the continuum limit is evaluated and then related to the Dirac equation. 
 
In this article, QW's and their relations to some known numerical schemes of the Dirac equation are reviewed from a slightly different perspective: it will be demonstrated that the most general QW's are obtained from lattice discretizations of the relativistic quantum wave equation for spin-1/2 particles. More precisely, starting from the continuum Dirac equation, it is shown that QW's can be placed in one-to-one correspondence with numerical schemes based on operator splitting and the QLB scheme. These numerical methods have been developed and employed as efficient numerical tools to solve relativistic quantum mechanics problems on classical computers \cite{FillionGourdeau20121403,QLB,FIL}. They have a number of interesting properties: they are easy to code, they can be easily parallellized and are very versatile. Moreover, their mathematical structure and the fact that the time discretization is realized by a set of unitary transformations makes the link to QW's possible. This connection is explored below and represents one of the main purposes of this article. 
Much of the material discussed here, in particular the numerical simulations, 
is not completely new, which is in line with the main purpose of this paper, namely 
an attempt to bridge the techniques utilized in numerical analysis and 
(quantum) Lattice Boltzmann theory to the language of quantum computing.

This article is organized as follows.
In Section \ref{sec:QW}, a general formulation of QW is presented, where the transfer matrix is time and space dependent. In Section \ref{sec:split}, the split operator method for the Dirac equation is presented, along with its exact correspondence with QW. Section \ref{sec:QLB} is devoted to the QLB method and connections with QW. Section \ref{sec:quantum_sim} is devoted to a qualitative discussion of the link between these numerical schemes and quantum computation. In Section \ref{sec:qe}, the schemes are casted in the form of a propagation-relaxation process and the notion of quantum equilibrium is introduced.
Based on the analogy between QW and QLB, a new QLB scheme for the (1+1) Dirac equation in curved space
is proposed in Section \ref{sec:QLBc}. Finally, the generalization of these methods to many dimensions is briefly discussed and numerical results are presented in Section \ref{sec:multiD}

\section{Quantum walks}
\label{sec:QW}

Let us consider a $(1+1)$ quantum walk on the line for a pair of complex wave functions (bi-spinor $\psi$),
obeying a discrete space-time evolution equations described by the following discrete map \cite{QW1,QW2,QW3,PhysRevA.77.032326}:
\begin{eqnarray}
\label{QW}
\begin{bmatrix}
\psi^{n+1}_{1,j} \\
\psi^{n+1}_{2,j}
\end{bmatrix}
= B_{j,n}
\begin{bmatrix}
\psi^{n}_{1,j-1} \\
\psi^{n}_{2,j+1}
\end{bmatrix}.
\end{eqnarray}
%
Here, the indices $j,m \in \mathbb{N} \otimes \mathbb{Z}$ label points on a discretization of space and time, respectively. The object $B_{j,n}$ is a two-by-two matrix with components \cite{QW3}
\begin{eqnarray}
B_{j,n}:= e^{-i \xi_{j,n}}
\begin{bmatrix}
e^{i\alpha_{j,n}} \cos\theta_{j,n} & e^{i \beta_{j,n}} \sin\theta_{j,n}\\
 -e^{-i \beta_{j,n}} \sin\theta_{j,n} & e^{-i\alpha_{j,n}} \cos\theta_{j,n}
\end{bmatrix}.
\end{eqnarray}
This matrix is a $U(2)$ operator ($B \in U(2)$) parametrized by the three space-time dependent
Euler angles $\theta_{j,n}, \alpha_{j,n}, \beta_{j,n}$ and a space-time dependent phase $\xi_{j,n}$. The latter is relevant when it depends on time and space, i.e. when it is local. If it is global ($\xi_{j,n}  = \xi$), it disappears from any observables and becomes unimportant. This occurs because when the phase is space-time dependent, it does not commute with the time and space translation operators. 

The matrix obeys $B \in SU(2)$ only when $\xi_{j,n} = k\pi$ for all $j,n$, with $k \in \mathbb{Z}$. Thus, this formulation is slightly more general than QW's considered in \cite{QW1,QW2,PhysRevA.77.032326} where $B \in SU(2)$ is studied. As seen in the next section, the choice $B \in U(2)$ will be important to have a general connection between mass terms of the Dirac equation and QW's. Finally, the $U(2)$ QW can also be implemented on quantum computers because the matrix $B$ is a unitary transformation: it represents the most general QW consistent with quantum computations.

In the above, the amplitudes $\psi_{1,2}$ code for the probability of the quantum
walker to move up (down) along the lattice site $j\in \mathbb{Z}$ at the time step $n \in \mathbb{N}$. This is a very rich structure, which has been shown to recover a variety of important quantum
wave equations, as soon as the Euler angles are allowed
to acquire a space-time dependence \cite{QW3}. In addition, it provides a wealth of potential algorithms for quantum computing. This was studied extensively in \cite{QW1,QW2,QW3} by analysing the continuum limit of these QW, yielding different versions of the Dirac equation. In this work, the opposite path is taken: it is shown that specific discretizations of the Dirac equation, using either a split-operator approach or the lattice Boltzmann technique, naturally lead to a QW formulation.

\section{Split-operator and quantum walks}
\label{sec:split}
The starting point of this discussion is the 1-D Dirac equation in Majorana representation written as (in units where $c=\hbar =1$):
\begin{eqnarray}
\label{eq:dirac}
i\partial_{t} \psi(z,t) = \left[ -i\sigma_{z} \partial_{z} + M(z,t) \right] \psi(z,t),
\end{eqnarray}
with the bi-spinor $\psi \in L_{2}(\mathbb{R},\mathbb{C}^{2})$. The generalized ``mass'' matrix $M$ is space and time dependent and may include contributions from the physical mass, the coupling to an electromagnetic potential or any other type of coupling. One requirement, however, is that $M$ is a Hermitian local operator without any derivatives. Generally, it can be written as
\begin{eqnarray}
M(z,t) &=&\mathbb{I}_{2}M_{0}(z,t) +  \boldsymbol{\sigma} \cdot \mathbf{M} ,\\
&=& \mathbb{I}_{2}M_{0}(z,t) +   \sigma_{x} M_{x}(z,t) + \sigma_{y} M_{y}(z,t) + \sigma_{z} M_{z}(z,t),
\end{eqnarray}
where $\mathbb{I}_{2}$ is the two-by-two identity matrix, $(\sigma_{i})_{i=x,y,z}$ are Pauli matrices and the coefficients $M_{0,x,y,z}$ represent the time- and space-dependent external fields which couple to the spinor.

The formal solution of Eq. \eqref{eq:dirac} is given by
\begin{eqnarray}
\label{eq:sol}
\psi(z,t) = \hat{T} \exp\left[ - \Delta t \sigma_{z} \partial_{z} - i \int_{t_{0}}^{t} M(z,t') dt'  \right] \psi(z,t_{0}),
\end{eqnarray}
where $\hat{T}$ is the time-ordering operator, $t_{0}$ is the initial time and $\Delta t = t-t_{0}$. Using an operator splitting technique, the solution in Eq. \eqref{eq:sol} can be approximated by \cite{FIL}
\begin{eqnarray}
\label{eq:sol_approx}
\psi(z,t) =  \exp\left[ - i \Delta t M(z,t_{0})  \right] \exp\left[ - \Delta t \sigma_{z} \partial_{z}   \right] \psi(z,t_{0}) + O(\Delta t^{2}).
\end{eqnarray}
The first exponential is a translation (streaming) operator which shifts the spinor components 
according to:
\begin{eqnarray}
\exp\left[ - \Delta t \sigma_{z} \partial_{z}   \right] \psi(z,t_{0}) =
\begin{bmatrix}
\psi_{1}(z-\Delta t, t_{0}) \\
\psi_{2}(z+\Delta t, t_{0})
\end{bmatrix}.
\end{eqnarray}
This suggests to use a spatial discretization where $\Delta z = \Delta t$ (this corresponds to a Courant-Friedrichs-Lewy (CFL) condition $C = c \Delta t/\Delta z = 1$, $c$ being the speed of light) such that the translation 
is exact on the lattice. 
Eq. \eqref{eq:sol_approx} can then be written as:
\begin{eqnarray}
\label{eq:dirac_dis}
\begin{bmatrix}
\psi^{n+1}_{1,j} \\
\psi^{n+1}_{2,j}
\end{bmatrix}
= \exp\left[ - i \Delta t M_{j,n}  \right]
\begin{bmatrix}
\psi^{n}_{1,j-1} \\
\psi^{n}_{2,j+1}
\end{bmatrix},
\end{eqnarray}
where we defined the following quantities on the lattice: $M_{j,n}:=M(n\Delta t, j\Delta z)$ and $\psi_{j}^{n}:=\psi(n\Delta t, j\Delta z)$. This last equation yields a numerical scheme to solve the Dirac equation. This numerical scheme has interesting properties, as discussed extensively in \cite{FIL,FillionGourdeau20121403,Lorin2011190}: it can be extended to higher order accuracy, it can be easily parallellized and it can be easily coded on a computer. In the following, it is also demonstrated that it is completely equivalent to the $U(2)$ QW described in the last section.

Eq. \eqref{eq:dirac_dis} is in the form of Eq. \eqref{QW}. Moreover, the exponential is also a unitary matrix:
\begin{eqnarray}
B' := \exp\left[ - i \Delta t M_{j,n}  \right] \in U(2),
\end{eqnarray}
and thus, there clearly exists a connection between $B$ and $B'$. They are expressed in different representation: $B$ uses the Euler angle parametrization while $B'$ is expressed in the canonical representation obtained by the exponential mapping of the Lie algebra. The latter is given explicitly by
\begin{eqnarray}
B' &=& \exp\left[ -i\Delta t M_{0,j,n} \right]\exp\left[ - i \Delta t \boldsymbol{\sigma} \cdot \mathbf{M}_{j,n}  \right] \\ 
&=&\exp\left[ -i\Delta t M_{0,j,n} \right] \left[ \mathbb{I}_{2} \cos(|\mathbf{M}_{j,n}|\Delta t) - i \cfrac{\boldsymbol{\sigma} \cdot \mathbf{M}_{j,n}}{|\mathbf{M}_{i,m}|}\sin(|\mathbf{M}_{j,n}|\Delta t) \right],
\end{eqnarray}
where
\begin{eqnarray}
|\mathbf{M}_{j,n}| = \sqrt{M_{x,j,n}^{2} + M_{y,j,n}^{2} + M_{z,j,n}^{2}}.
\end{eqnarray}
It is well-known that parametrizations of $U(2)$ matrices are related to each other \cite{gilmore2012lie} and it can be determined that the mapping between both representations is given by
\begin{eqnarray}
\label{eq:rel_split1}
\xi_{j,n} &=& M_{0,j,n} \Delta t ,\\
\label{eq:rel_split2}
\tan (\alpha_{j,n}) &=& -\frac{M_{z,j,n}}{|\mathbf{M}_{j,n}|} \tan(|\mathbf{M}_{j,n}| \Delta t) ,\\
\label{eq:rel_split3}
\tan (\beta_{j,n}) &=& \frac{M_{x,j,n}}{M_{y,j,n}}, \\
\label{eq:rel_split4}
\tan (\theta_{j,n}) &=& \tan(|\mathbf{M}_{j,n}| \Delta t) \sqrt{\frac{M_{x,j,n}^{2} + M_{y,j,n}^{2}}{|\mathbf{M}_{j,n}|^{2} + M_{z,j,n}^{2} \tan^{2}(|\mathbf{M}_{j,n}|\Delta t)}}.
\end{eqnarray}
When $M_{0} = 0$, one recovers the $SU(2)$ QW because then, $\tilde{B} := B'|_{M_{0} = 0} \in SU(2)$. These last equations give a one-to-one correspondence between the QW formulation, characterized by the parameters $\alpha_{j,n},\beta_{j,n},\theta_{j,n},\xi_{j,n}$, and the discretization of the Dirac equation with a generalized mass term $M_{j,n}$. Therefore, the last results show that in the continuum limit, every space-time dependent QW on the line becomes a time-dependent 1D Dirac equation with a specific mass matrix. A given QW can thus be fully characterized by the relativistic dynamics of an electron coupled to a space-time dependent external field. This occurs because there is an equivalence between the discretization of the Dirac equation, based on operator splitting, and the QW formulation. Moreover, this connection may be the base for the implementation of a quantum algorithm that solves the Dirac equation on quantum computers.

The operator splitting technique presented here also bears a close relationship with the QLB technique, to which we now turn.

\section{Quantum lattice Boltzmann, operator splitting and Quantum Walks}
\label{sec:QLB}
The QLB was inspired by a direct analogy between the way the Dirac equation goes
to the Schroedinger equation in the limit $v/c \to 0$, and the way that the Navier-Stokes
equations of classical fluid-dynamics emerge from the Boltzmann equation in the limit
of small Knudsen number, $Kn \to 0$, where $Kn=l/L$ is the ratio of the
molecular mean free path to the typical macroscopic scale.
In both cases, the smallness parameter controls the enslaving of the fast modes
to the slow ones: {\it non-equilibrium} to {\it equilibrium} for the classical
case, versus {\it excited} states to {\it ground} state in the quantum one.
Of course, the quantum case shows no genuine relaxation since its dynamics is reversible.
Yet, enslaving can be interpreted in the sense of fast oscillations around a local quantum
equilibrium (Zitterbewegung), which average out once time is coarse-grained on a scale
larger than the period of the fast oscillations. So, Zitterbewegung may be regarded as the quantum relativistic analogue of
classical non-equilibrium fluctuations.

Based on this analogy, QLB was formulated as a lattice Boltzmann analogue
of the Dirac equation in the Majorana representation, where the streaming matrix is {\it real}. To obtain the QLB scheme, it is convenient to write the Dirac equation as
\begin{eqnarray}
\left[ \partial_{t} + v_{a}\partial_{z} \right]\psi_{a}(z,t) = -i\sum_{b}M_{ab}(z,t) \psi_{b}(z,t),
\end{eqnarray}
where $a,b$ are spinor indices and the ``microscopic velocities'' are given by $v_{1,2} = \pm 1$. This is clearly in a ``Boltzmann-like'' form with two discrete velocities and a collision term $M$. When QLB is employed in fluid mechanics to solve the continuum equations of motion, there is a family of possible numbers of discrete velocities for a given lattice \cite{0295-5075-17-6-001} and each choice yields a different numerical scheme (for instance, the 9 velocities scheme in 2-D and the 27 velocities scheme in 3-D are popular choices on square lattices \cite{PhysRevE.56.6811}). For the Dirac equation, this choice is dictated by the mathematical structure of the equation.

The lattice Boltzmann equation is then written as \cite{succi2001lattice,chen1998lattice} 
\begin{eqnarray}
\label{eq:LB_scheme_no}
\psi_{i}(z + v_{a} \Delta t,t + \Delta t) = \psi_{a}(z,t) - i\Delta t\sum_{b}M_{ab}(z,t) \psi_{b}(z,t) + O(\Delta t^{2}).
\end{eqnarray}
As for the splitting method described in the last section, this suggests to use a space discretization where $\Delta z = \Delta t$. Using a ``naive'' approach in line with LB methods in fluid mechanics, the last equation would become
\begin{eqnarray}
\label{eq:LB_scheme_dis}
\begin{bmatrix}
\psi_{1,j+1}^{n+1} \\
\psi_{2,j-1}^{n+1}
\end{bmatrix}
 = 
\begin{bmatrix}
 \psi_{1,j}^{n} \\
 \psi_{2,j}^{n}
\end{bmatrix} 
- i\Delta tM_{j,n} 
\begin{bmatrix}
 \psi_{1,j}^{n} \\
 \psi_{2,j}^{n}
\end{bmatrix}.
\end{eqnarray}
Then, the matrix $-iM$ on the right acts as collision operator while the streaming is executed by the left part of the equation. This scheme is derived by using the formal analogy between the Dirac equation, the Boltzmann equation and the LB technique. However, the resulting numerical method is unstable and the $L^{2}$ norm is not preserved \cite{succi2002kinetic}. 

It is possible, however, to recover a stable and norm-preserving method. Instead of using the ``naive'' LB equation discretization given in Eq. \eqref{eq:LB_scheme_dis}, the mass term on the right-hand side of Eq. \eqref{eq:LB_scheme_no} is discretized by using an implicit Crank-Nicolson average:
\begin{eqnarray}
iM(z,t)
\begin{bmatrix}
 \psi_{1}(z,t) \\
 \psi_{2}(z,t)
 \end{bmatrix} = \cfrac{i}{2} M_{j,n}
 \left\{ 
\begin{bmatrix}
\psi_{1,j+1}^{n+1} \\
 \psi_{2,j-1}^{n+1}
 \end{bmatrix}  
 +
 \begin{bmatrix}
\psi_{1,j}^{n} \\
 \psi_{2,j}^{n}
 \end{bmatrix}  
  \right\} .
\end{eqnarray} 
Reporting this into Eq. \eqref{eq:LB_scheme_no}, one obtains the second order accurate QLB scheme:
\begin{eqnarray}
\label{eq:qlb}
\begin{bmatrix}
\psi^{n+1}_{1,j+1} \\
\psi^{n+1}_{2,j-1}
\end{bmatrix}
= T_{j,n}
\begin{bmatrix}
\psi^{n}_{1,j} \\
\psi^{n}_{2,j}
\end{bmatrix} + O(\Delta t^{2}),
\end{eqnarray}
where the transfer matrix is given by
\begin{eqnarray}
T_{j,n} = \left[\mathbb{I}_{2} + i \cfrac{\Delta t}{2} M_{j,n}  \right]^{-1} \left[\mathbb{I}_{2} -  i \cfrac{\Delta t}{2}M_{j,n}\right].
\end{eqnarray}
To link these results with the ones of the last sections, it is convenient to shift the spinor components by one lattice point (we let $\psi_{1,j+1} \rightarrow \psi_{1,j}$ and $\psi_{2,j-1} \rightarrow \psi_{2,j}$). Then, the QLB scheme becomes
\begin{eqnarray}
\begin{bmatrix}
\psi^{n+1}_{1,j} \\
\psi^{n+1}_{2,j}
\end{bmatrix}
= T_{j,n}
\begin{bmatrix}
\psi^{n}_{1,j-1} \\
\psi^{n}_{2,j+1}
\end{bmatrix} + O(\Delta t^{2}),
\end{eqnarray}
which is in the same form as the QW and the splitting method presented previously. The transfer matrix can be evaluated explicitly. It is given by
\begin{eqnarray}
\label{eq:coll_mat_qlb}
T_{j,n} = \frac{1}{C_{j,n}}
\begin{bmatrix}
1 - i\Delta t M_{z,j,n} + \frac{\Delta t^{2}}{4} M_{j,n}^{2} & -\Delta t \left(iM_{x,j,n} + M_{y,j,n}  \right)\\
-\Delta t \left(iM_{x,j,n} - M_{y,j,n}  \right) & 1 + i\Delta t M_{z,j,n} + \frac{\Delta t^{2}}{4} M_{j,n}^{2}
\end{bmatrix},
\end{eqnarray} 
where $C_{j,n} = 1 + i\Delta t M_{0,j,n} - \frac{\Delta t^{2}}{4} M_{j,n}^{2}$ and $M_{j,n}^{2} = M_{0,j,n}^{2} - \mathbf{M}_{j,n}^{2}$. It can be readily verified that $T_{j,n}T_{j,n}^{\dagger} = \mathbb{I}_{2}$. As a consequence, the transfer matrix is unitary $T_{j,n} \in U(2)$, for any size of the time step and thus, conserves the $L^{2}$ norm. Moreover, just like the splitting method, there exists a correspondence with QW's because both have $U(2)$ collision matrices. The identification with the general QW in Eq. (\ref{QW}) yields
\begin{eqnarray}
\tan(\xi_{j,n}) &=& \cfrac{\mathrm{Im} (C_{j,n})}{\mathrm{Re} (C_{j,n})} = \cfrac{ \Delta t M_{0,j,n} }{1 - \frac{\Delta t^{2}}{4} M_{j,n}^{2}} \\
\tan(\alpha_{j,n}) &=&- \cfrac{\Delta t M_{z,j,n}}{1 + \frac{\Delta t^{2}}{4} M_{j,n}^{2}}, \\
\tan(\beta_{j,n}) &=& \cfrac{M_{x,j,n}}{M_{y,j,n}} ,\\
\tan(\theta_{j,n}) &=& - \Delta t \sqrt{\cfrac{M_{x,j,n}^{2}+M_{y,j,n}^{2}}{\left(1+ \frac{\Delta t^{2}}{4} M_{j,n}^{2}   \right)^{2} + \Delta t^{2} M_{z,j,n}^{2}}}.
\end{eqnarray}
These relations give a correspondence between the QLB technique and the QW. They are similar to the ones for the splitting method displayed in Eqs. \eqref{eq:rel_split1} to \eqref{eq:rel_split4} and actually serve the same purpose: they allow to map a numerical scheme to the QW. The differences are obviously due to the fact that QLB is based on a LB formulation combined with a Crank-Nicolson average to insure stability while the splitting scheme separates the Dirac equation into different operators which can be integrated exactly. However, both methods share the same general structure where a streaming step is followed by a collision step.

\subsection{An explicit example: the free case}

The Dirac equation for the massive and free case (no external field coupling) is 
\begin{eqnarray}
\label{eq:dirac_maj}
i\partial_{t} \psi(z,t) = \left[ -i\sigma_{z} \partial_{z} + \sigma_{y} m \right] \psi(z,t),
\end{eqnarray}
where $M_{y} = m$ is a physical fermion mass. 
This representation of the Dirac equation yields real spinor components, as readily seen by writing the equation componentwise:
\begin{eqnarray}
\label{MAIO}
(\partial_{t}  + \partial_z) \psi_{1}(z,t) = -m \psi_{2}(z,t),\\
(\partial_t  -  \partial_z )\psi_{2}(z,t) = +m \psi_{1}(z,t).
\end{eqnarray}
The QLB scheme, for the massive and free case, reads as follows \cite{QLB}:
\begin{eqnarray}
\begin{bmatrix}
\psi^{n+1}_{1,j} \\
\psi^{n+1}_{2,j}
\end{bmatrix}
= T_{\rm free}
\begin{bmatrix}
\psi^{n}_{1,j-1} \\
\psi^{n}_{2,j+1}
\end{bmatrix},
\end{eqnarray}
where the transfer matrix is can be obtained from Eq. \eqref{eq:coll_mat_qlb} by setting $M_{0} = M_{x} = M_{z} = 0$ and $M_{y} = m$: 
\begin{eqnarray}
T_{\rm free}:=
\begin{bmatrix}
a(m) & b(m)\\
 -b(m) & a(m)
\end{bmatrix} = \cfrac{1}{1+\frac{\Delta t^{2}m^{2}}{4}} 
\begin{bmatrix}
1-\cfrac{\Delta t^{2}m^{2}}{4} &- m\Delta t \\
m \Delta t  & 1-\cfrac{\Delta t^{2}m^{2}}{4}
\end{bmatrix},
\end{eqnarray}
It is readily seen that $a(m)$ and $b(m)$ are second-order Pade'-like
approximants of $\cos(m)$ and $\sin(m)$, respectively.
This is the natural consequence of the implicit time-marching (Crank-Nicolson) scheme
as applied to the (1+1) Dirac equation in Majorana form.
For a free particle, the correspondence to QW is particularly simple and is given by 
\begin{eqnarray}
\label{eq:map_QW1}
\xi = \alpha=\beta=0, \\
\label{eq:map_QW2}
\tan(\theta) =- \frac{m\Delta t}{1-\frac{\Delta t^{2}m^{2}}{4}}.
\end{eqnarray}
This mapping of the QLB to the QW is exact for any value of $m$.

\section{Prospects for quantum simulation}
\label{sec:quantum_sim}

The numerical methods described in Sections \ref{sec:split}-\ref{sec:QLB} 
can be straightforwardly implemented on classical computers. 
However, the stream-collide structure of these numerical scheme makes them 
suitable for an efficient implementation on quantum computers as well. 
In particular, they can be written as:
\begin{eqnarray}
\label{QW_gen}
\psi^{n+1}_{j}  
= B_{j,n} S_{j} \psi^{n}_{j},
\end{eqnarray}
where $S_{j}$ is the shift operator, defined as:
\begin{eqnarray}
\label{shift_op}
S_{j} \psi^{n}_{j}
= 
\begin{bmatrix}
\psi^{n}_{1,j-1} \\
\psi^{n}_{2,j+1}
\end{bmatrix}.
\end{eqnarray}
This shift operator is a unitary operation and it can be realized experimentally 
by using fundamental quantum gates \cite{PhysRevA.72.062317,PhysRevA.77.032326}. 
The rotation operator $B_{j,n}$ belongs to $U(2)$ and therefore can also 
be realized by these quantum gates, as any other unitary transformations \cite{kaye2006introduction}. 
Therefore, it is possible to map the numerical method to quantum walk, which can be implemented 
efficiently on quantum computers. This mapping is possible because each step of the scheme 
is a unitary transformation: this makes these schemes norm-preserving 
and sets the link with quantum computations. 

The latter would be particularly useful for the study of 
relativistic quantum systems where a time-dependent solution of the Dirac equation 
is required, such as in very high intensity laser physics \cite{RevModPhys.84.1177} or 
graphene physics \cite{geim2007rise}.

Another subject of major interest for future research is the extension of 
the QLB methodology to quantum many body systems and quantum field 
theory, two paramount sectors of modern physics which are particularly exposed 
to the limitations of classical (non-quantum) electronic computing. 
Progress in this direction depends on the ability to replace the quantum wavefunction
by the corresponding second-quantized quantum operators, and show that the dynamics
of the second-quantized QLB scheme still preserves the appropriate 
equal-time commutation relations. 
Preliminary efforts along this line have been developed in \cite{1751-8121-40-26-F07}
in $1+1$ dimensions. Extensions to strongly non-linear field theories in $d>1$
remain to be explored.
As to quantum-many body problems, LB-like methods have been recently
adapted to electronic structure simulations \cite{PhysRevLett.113.096402} 
In this work, a classical LB scheme is employed to solve the Kohn-Sham equations of 
density functional theory in the form diffusion-reaction equations in imaginary time.
Allied QLB schemes could prove very useful to solve the corresponding real-time
quantum many-body transport problems within the framework of
time-dependent density functional theory.

Finally, we wish to point out the intriguing possibility of realizing both quantum
and classical LB schemes on quantum analogue simulators, as recently explored in \cite{mezzacapo2015quantum}.

\section{Quantum equilibria}
\label{sec:qe}

In view of quantum computing implementations, it is of interest
to cast the Dirac equation in the form of a propagation-relaxation process, where
the collision matrix is now interpreted as a scattering process, relaxing the spinor component
around a local quantum equilibrium. 

For this purpose, it is useful to reconsider the 1-D Dirac equation in the form:  
\begin{eqnarray}
\left[ \partial_{t} + v_{a}\partial_{z} \right]\psi_{a}(z,t) = -i\sum_{b}M_{ab}(z,t) \psi_{b}(z,t).
\end{eqnarray}
Then, it is formally possible to  define a local equilibrium as:
\begin{eqnarray}
\label{LEQ}
\psi_{\rm eq}(z,t) :=U\psi(z,t),
\end{eqnarray}
where $U$ is a unitary matrix that depends on $M$. This transformation is chosen to recast the Dirac equation in relaxation form:
\begin{eqnarray}
\left[ \partial_{t} + \sigma_{z}\partial_{z} \right]\psi(z,t) = 
- \Omega \left[ \psi(z,t) - \psi_{\rm eq}(z,t)  \right],
\end{eqnarray}
where $\Omega = iM [\mathbb{I} + U]^{-1}$. The explicit value of $U,\Omega$ is not unique but a convenient choice is $U = e^{iM \tau}$.

In this vests, the Dirac equation looks formally like a linear Boltzmann equation 
for two-component models in the single relaxation time approximation \cite{succi2001lattice}. 
Therefore, it can be interpreted as a propagation-relaxation process in imaginary time, 
whereby collisions, implemented by the scattering operator $\Omega$, drive oscillations
around the equilibrium distribution $\psi_{\rm eq}$. 
By defining a post-collision wavefunction as:
\begin{equation}
\label{POST}
\psi^{'}(z,t) := \left( 1 - \Delta t \Omega  \right) \psi(z,t)  +  \Delta t \Omega  \psi_{\rm eq}(z,t),
\end{equation}
the Dirac equation takes the most compact form
\begin{equation}
\label{QCOMP}
\psi(z+v_a \Delta t, t+\Delta t)  = \psi^{'}(z,t).
\end{equation}
This is particularly suitable to quantum computing implementations in the
form of a classical stream-collide dynamics.
The collision (relaxation) gate transforms the pre-collisional spinor
$\psi$ into the post-collisional state $\psi'$, and the streaming gate moves
the post-collisional spinor to its destination location $z \pm  \Delta z$.
Both operations are unitary and can be encoded in logical gates 
for quantum computing purposes \cite{MEY,BOG}.

The expression (\ref{LEQ}) shows that the local equilibria is a linear function
of the actual wavefunction $\psi$, hence itself a function of space and time.

The question is: will the actual wavefunctions ever reach this moving target, i.e,
$\psi = \psi_{\rm eq}$?

Based on its definition, this can only occur once $\psi_{\rm eq}$ lies in the null-space 
of the scattering matrix $M$, namely:
\begin{equation}
\label{EQUIL}
M \psi_{\rm eq} = 0.
\end{equation}
For the case of a free massive particle, it can be checked explicitly
that the only solution is the trivial vacuum $\psi_{\rm eq} \equiv 0$. 

This means that the spinorial wavefunction is a superposition of (both slow and fast) 
zero-average oscillations around a local equilibrium which, consistently
with the reversible nature of quantum mechanics, is actually never attained.

Although this remains to be checked in detail, we conjecture that
the same holds true for the case of a massive particle in an external 
potential, because in this case the Dirac equation is still linear. 

Based on (\ref{EQUIL}), the condition for the local equilibrium
to depart from a trivial vacuum is that the matrix $M$ be singular,
i.e. the local equilibrium is a zero-mode of the scattering matrix.

Non-trivial quantum zero-modes, $\psi_{\rm eq} \neq 0$, may indeed arise for {\it non-linear} quantum wave 
equations, such as the Gross-Pitaevski or the mean-field version of 
the Nambu-Jona-Lasinio model, to be dicussed shortly. 
A non-trivial local equilibrium would then signal a spontaneously
broken symmetry, which is indeed the distinctive trait of the 
aforementioned non-linear quantum wave equations. 

Even though the notion of quantum equilibrium remains purely formal
in nature, it is argued that it might nonetheless facilitate 
quantum computing implementations based on the compact expressions (\ref{POST})
and (\ref{QCOMP}). This stands out a very interesting topic for future research.

\section{QLB in curved space-time}
\label{sec:QLBc}

Quantum walks have been shown to map into Dirac-like equations in curved space as well by evaluating the continuum limit of certain QW's \cite{PhysRevA.88.042301,QW3}. Here, in the same spirit as other sections, a QW structure is obtained by discretizing the Dirac equation in curved space time using a QLB-like approach. However, because the wave function propagates on a curved manifold, the structure of the resulting QW is different from Eq. \eqref{QW} and should include a residency matrix that corrects the streaming step, which is strictly valid only in flat space. This is different from the result obtained in \cite{PhysRevA.88.042301,QW3}.

%
The Dirac equation in a static (1+1) curved space writes as:
\begin{equation}
\label{DIRACG}
\gamma^a [e^{\mu}_a (\partial_{\mu} -i A_{\mu})]\psi +
\frac{1}{2}g^{-1/2} \partial_{\mu} (g^{1/2} e^{\mu}_a) \psi)] =-i m \psi,
\end{equation}
where $a=0,1$ and $\mu=t,z$.
In the above $e^{\mu}_a$ is the two-dimensional vierbein (zweibein) relating
the components of the locally tangent Minkowski space basis $(e_0,e_1)$, described by the metric $\eta_{ab}$, to the global space basis $(e_t,e_z)$, described by the general metric $g_{\mu \nu}$.  Also, $g:= \mathrm{det} (g_{\mu \nu})$ is the determinant of the metric tensor.

The general form of the corresponding partial differential equation is
\begin{equation}
\partial_t \psi + \sigma_{z}A(z)\partial_z \psi = Q(z,t) \psi,
\end{equation}
with $A(z) \in \mathbb{R}$ a function of space and $Q(z,t)$ the two-by-two gravitational collision matrix
associated with Eq. (\ref{DIRACG}). 
This becomes a hyperbolic system of equations where the advection
speed $A(z)$  should not be confused with the vector electrodynamic potential.
Since the advection term is heterogeneous, a strict QLB structure, i.e. streaming
along constant lightcones $\Delta z= \mp c \Delta t$, is no longer viable.
Here, the situation is very similar to classical LB schemes on non-uniform grids
(unsurprisingly, since in dimension $D=1$, gravity is basically a stretching of the metric).
For this purpose, several finite-volume LB (FVLB) schemes have been formulated,
whose main outcome is that the streaming operator is no longer a diagonal matrix
with eigenvalues $\pm 1$, as required by the formal QLB structure \cite{FVLB,nannelli1992lattice}.

\begin{figure}
\includegraphics[scale=0.5]{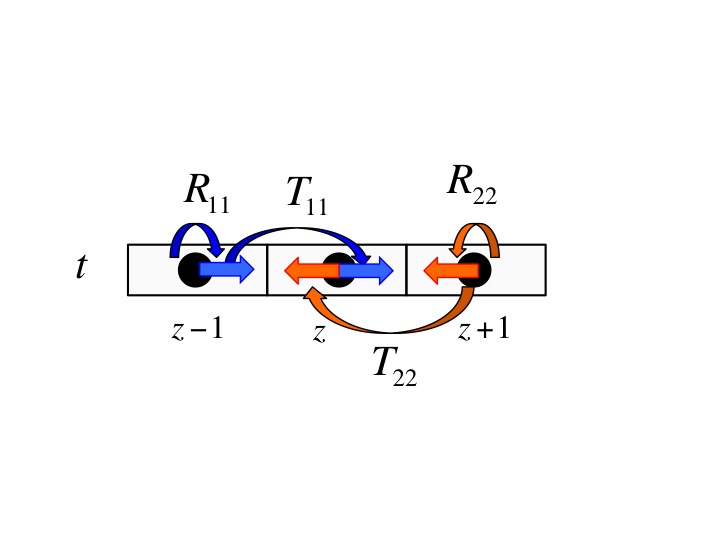}
\caption{Sketch of the transfer and residency matrix elements.
$T_{11}$ is the fraction of up-moving spinor jumping from $j-1$ at time $n-1$
to $j$ at time $n$, while $R_{11}$ is the fraction left in $j-1$.
$T_{22}$ and $R_{22}$ bear the same meaning for the down moving spinor.
}
\label{fig:trans_res}
\end{figure}

In this respect, a finite-volume QLB scheme for the Dirac equation in curved space can be obtained by writing the Dirac equation as
\begin{equation}
\partial_t \psi +\partial_z \left[ \sigma_{z}A(z) \psi \right] = Q'(z,t) \psi,
\end{equation}
with $Q'(z,t) := Q(z,t)+\partial_{z} A(z)\sigma_{z}$. Then, this equation is integrated
over a control volume $\mathcal{V}$ (enclosed by a surface $\mathcal{S}$), extending from $(j-\frac{1}{2})\Delta z$ to $(j+\frac{1}{2})\Delta z$. Finally, applying a Crank-Nicolson time-marching and combining with
upwind finite-differences, the discretized equation is given by
\begin{eqnarray}
\psi_{1,j}^{n+1}-\psi_{1,j}^{n} + \frac{1}{c}\oint A\psi_{1} d\mathcal{S} &=& \frac{1}{2c} (Q'_{11,j-1}\psi_{1,j-1}^{n}+Q'_{11,j}\psi_{1,j}^{n+1})  \nonumber \\
                                    && + \frac{1}{2c}  (Q'_{12,j+1}\psi_{2,j+1}^{n}+Q'_{12,j}\psi_{2,j}^{n+1}),\\
\psi_{2,j}^{n+1}-\psi_{2,j}^{n} - \frac{1}{c}\oint A\psi_{2} d\mathcal{S} &=& \frac{1}{2c} (Q'_{22,j+1}\psi_{2,j+1}^{n}+Q'_{22,j}\psi_{2,j}^{n+1}) \nonumber \\
                                    && + \frac{1}{2c}  (Q'_{21,j-1}\psi_{1,j-1}^{n}+Q'_{21,j}\psi_{1,j}^{n+1}),
\end{eqnarray}
where $c := \Delta z/\Delta t$ is the uniform lattice light speed (or CFL condition).
The boundary integrals are given by
\begin{eqnarray}
\oint A\psi_{1,2} d\mathcal{S} = [A \psi_{1,2}]_{j+\frac{1}{2}} - [A \psi_{1,2}]_{j-\frac{1}{2}},
\end{eqnarray}
an thus, require an interpolation from the cell centers $j$, to the
north and south boundaries at $j \pm 1/2$, respectively.
Following a common practice in finite-volume formulations of hyperbolic problems, the flux terms are approximated by \cite{leveque2002finite}:
\begin{eqnarray}
\left[A \psi_{1}\right]_{j+\frac{1}{2}} &=& A_{j} \psi_{1,j}^{n}, \\
\left[A \psi_{1}\right]_{j-\frac{1}{2}} &=& A_{j-1} \psi_{1,j-1}^{n} ,\\
\left[A \psi_{2}\right]_{j+\frac{1}{2}} &=& A_{j+1} \psi_{2,j+1}^{n}, \\
\left[A \psi_{2}\right]_{j-\frac{1}{2}} &=& A_{j} \psi_{2,j}^{n} .
\end{eqnarray}
%
%
%
As a result:
%
%
\begin{eqnarray}
\left(1- \frac{Q'_{11,j}}{2c}  \right)\psi_{1,j}^{n+1} - \frac{Q'_{12,j}}{2c}\psi_{2,j}^{n+1} &=& 
\left(1 - \frac{A_{j}}{c}  \right) \psi_{1,j}^{n} 
+\left(\frac{Q'_{11,j-1}}{2c} +  \frac{A_{j-1}}{c} \right)   \psi_{1,j-1}^{n}  \nonumber \\
                                    && + \frac{Q'_{12,j+1}}{2c}  \psi_{2,j+1}^{n},\\
\left(1- \frac{Q'_{22,j}}{2c}  \right)\psi_{2,j}^{n+1} - \frac{Q'_{21,j}}{2c}\psi_{1,j}^{n+1} &=& 
\left(1 - \frac{A_{j}}{c}  \right) \psi_{2,j}^{n} 
+\left(\frac{Q'_{22,j-1}}{2c} +  \frac{A_{j+1}}{c} \right)   \psi_{2,j-1}^{n}  \nonumber \\
                                    && + \frac{Q'_{21,j+1}}{2c}  \psi_{1,j+1}^{n}.
\end{eqnarray}
%
%
%
It is readily seen that this reduces to a standard QLB in the limit of a uniform
grid, when $A_{j}/c =1$. In this case the streaming is diagonal with speed $\pm c$
and the spinors at $(j,n+1)$ are connected to the corresponding
spinors at $(j \pm 1,n)$ by a local
$2 \times 2$ matrix, which can be readily inverted to deliver a fully explicit map.
However, when $A_{j}/c \ne 1$, the spinors at $(j,n)$ also enter the map, so that local
inversion delivers a slightly more elaborated structure, namely:
\begin{eqnarray}
\psi^{n+1}_{j} = (R + TS)\psi^{n}_{j} ,
\end{eqnarray}
which can be written more explicitly as
\begin{eqnarray}
\label{GQLB}
\begin{bmatrix}
\psi^{n+1}_{1,j} \\
\psi^{n+1}_{2,j}
\end{bmatrix}
= R
\begin{bmatrix}
\psi^{n}_{1,j} \\
\psi^{n}_{2,j}
\end{bmatrix}
+
 T
\begin{bmatrix}
\psi^{n}_{1,j-1} \\
\psi^{n}_{2,j+1}
\end{bmatrix}.
\end{eqnarray}
%
In the above $T$ is the local $2 \times 2$ transfer matrix including collisions, $S$ is the streaming operator and $R$ the local residency matrix, expressing the fraction
of spinors which are left in the cell centered about $z$ as the quantum system advances from $t$ to $t+\Delta t$. This is depicted in Fig. \ref{fig:trans_res}.

Clearly, the residency matrix vanishes in the case of a uniform mesh, i.e. no gravity.
The mapping (\ref{GQLB}) represents the ``gravitational'' QLB.
The detailed expressions of the streaming and residency matrices depend on the specific
form of the metric tensor and associated vierbeins. Moreover, this analysis concentrates on the mathematical structure (streaming and collision steps) of the resulting scheme rather than on its numerical properties (convergence, stability, etc). These topics shall make the object
of a future publication.

\section{Multi-dimensions}
\label{sec:multiD}

The discretization presented in this work extends to the $D+1$ dimensional case by applying the notion of operator splitting.
This implies the inclusion of a new dynamic step which is entirely quantum: namely a
"rotation", designed so as to keep the spin aligned with the momentum along
each of the three spatial directions. Schemes using this strategy can be found in \cite{QLB,FIL} and in \cite{FillionGourdeau20121403} for higher order splittings. 

It might be that such rotation is not needed by formulating the Dirac equation as a random walk on
other lattice with more natural topologies (the diamond) lattice.
The QLB-QW equivalence in multi-dimensions will be discussed in a future publication. However, to demonstrate the strength of the numerical schemes presented here and to show some possible applications for quantum computing, numerical results in 2-D are presented in the following. 

\subsection{Numerical results}

As an example of possible applications of QLB scheme, we present
two representative simulations: Klein tunnelling in the presence of
random impurities and Dirac equation
with Nambu-Jona-Lasinio (NJL) interactions in $2+1$ space-time dimensions. 
Details on the numerical methods used to obtain these results are given in \cite{QLB,FIL,nonlinear_dirac}. Also, these results are not completely new as similar systems have been studied in \cite{FIL,RANDOM}.

\subsubsection{Graphene with random impurities}

\begin{figure}
  \centering
  \includegraphics[scale=0.20]{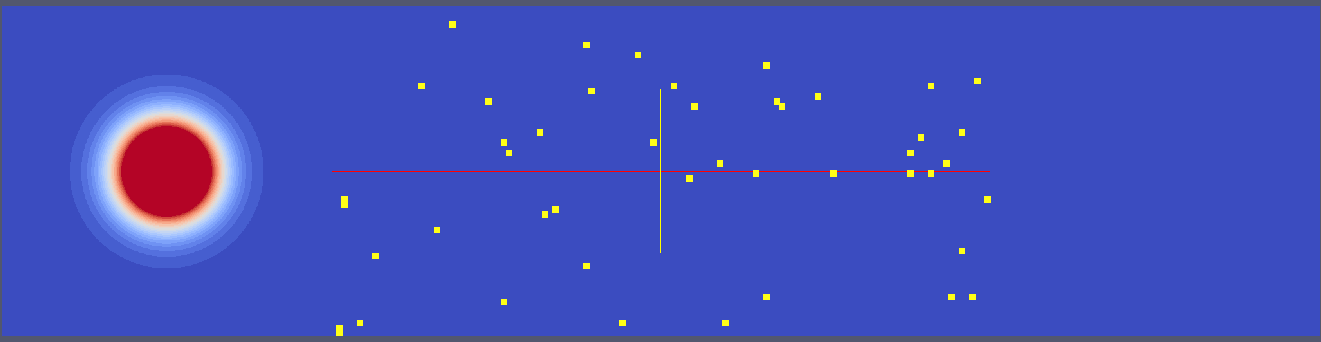} \\ \vspace{0.2cm}
  \includegraphics[scale=0.20]{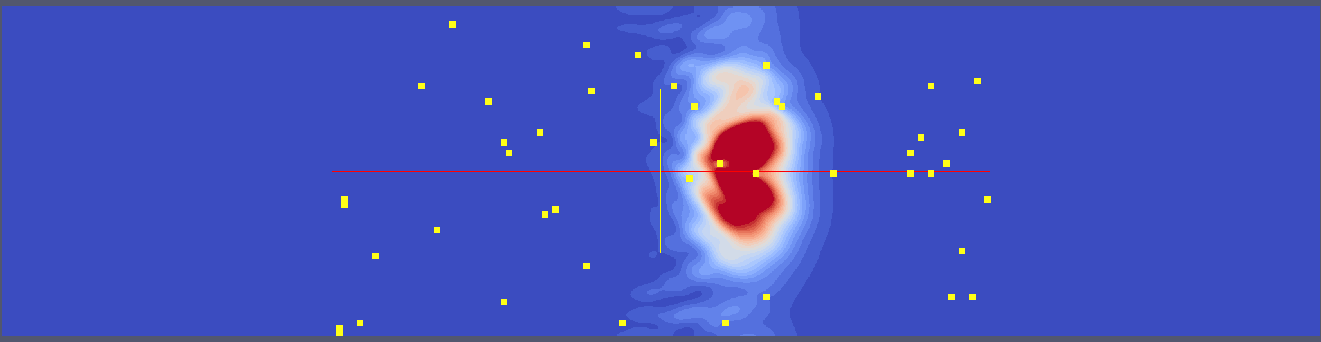} \\ \vspace{0.2cm}
  \includegraphics[scale=0.20]{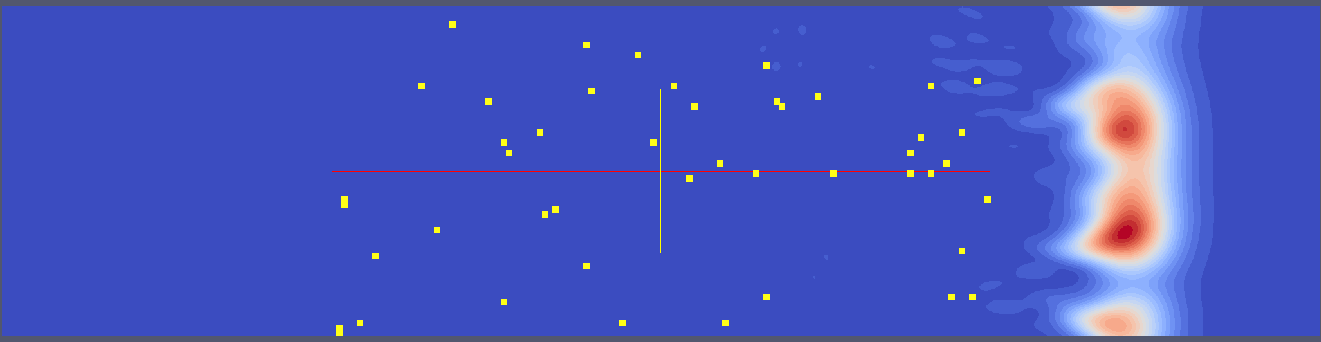} \\ \vspace{0.2cm}
  \includegraphics[scale=0.20]{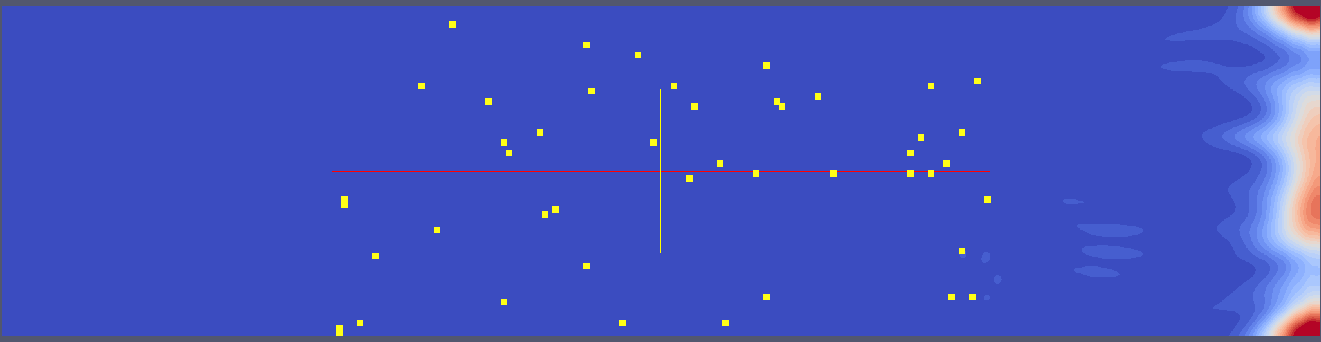}
  \caption{Wave packet density $\rho$ at times = $0$, $900$, $1500$,
    and $1800$ (lattice units) for a Gaussian wave packet propagating through a
    graphene sample with randomly distributed impurities. The simulation is performed with
    impurity percentage $C=0.5\%$ and impurity potential $V = 50$ MeV.}
  \label{fig:imp05_v50}
\end{figure}

In the first numerical test, the propagation of a Gaussian wave packet through a graphene sample
 with randomly distributed impurities is simulated \cite{RANDOM}.
In Ref. \cite{RANDOM}, simulations are performed for
different  values of the impurity concentration and the potential
barrier, in order to provide an estimate of the effect
of impurity concentration on the conductivity of the graphene sample. 
In \figref{imp05_v50}, we report some representative snapshots of the first
$1800$ time steps of the simulation, at an impurity percentage$=0.5\%$
and $V=50$ MeV.
A lattice of size $2048\times 512$ cells is used and the cell size is chosen 
to be $\Delta z = 0.96$ nm, while the spreading
of the initial Gaussian wave packet is $\sigma = 48$ (in numerical
units), leading to a Fermi frequency $k_F = 0.117$ ($80$ MeV in
physical units). In this simulation, a fully relativistic particle
($m=0$) is considered.

From \figref{imp05_v50}, we can see that the wave packet is scattered by
the impurities, giving rise to a plane front out of  the initial
Gaussian configuration. As a consequence of the randomness induced in the
wave function by the disordered media, there is a momentum loss
and therefore the motion of the wave packet is found
to experience a corresponding slow down. 
It is also found that the wave packet takes more time to regroup as the impurity 
concentration and impurity potential are increased.

\subsubsection{Nambu-Jona-Lasinio interaction}

As a second example, we present a $2+1$ space-time simulation of the
Dirac equation with a Nambu-Jona-Lasinio (NJL) interaction \cite{NJL1}. The Dirac equation with a NJL
interaction term driven by the coupling parameter $g$, reads as follows
\begin{equation}
\label{eq:DiracNJL}
(i \gamma^{\mu} \partial_{\mu} -m ) \psi + g ( (\bar \psi \psi) \psi +
(\bar \psi \gamma^5 \psi) \gamma^5 \psi) = 0,
\end{equation}
where $\psi = (\psi_1, \psi_2, \psi_3, \psi_4)^T$, $\gamma^5
\equiv i \gamma^0 \gamma^1 \gamma^2 \gamma^3$ and $\bar \psi =
\psi^{\dag} \gamma^0$.

This model represents a paradigm for dynamic mass acquisition via spontaneous symmetry breaking
due to the non-linear interactions.

Let us consider an initial condition
given by the following Gaussian minimum-uncertainty
wave packet:
\begin{equation}
G_0(z,y) = (2 \pi \sigma^2)^{-1/2} \exp(-\frac{z^2+y^2}{4 \sigma^2}),
\end{equation}
centered about $(z,y)=(0,0)$, with initial width $\sigma$.
Let $k_z$ and $k_y$ be the initial energy of the wave packet
and impose the following initial condition:
\begin{equation}
\label{eq:IC_2d}
\begin{split}
u_1(z,y) = u_2(z,y)  &= C_u G_0(z,y) \exp(i(k_z z + k_y y)), \\
d_1(z,y) = d_2(z,y)  &= C_d G_0(z,y) \exp(-i(k_z z + k_y y)),
\end{split}
\end{equation}
where coefficients $C_u$ and $C_d$ obey the condition
$2 C_u^2 + 2 C_d^2 = 1 $,
so that $\rho = |\psi_1|^2 + |\psi_2|^2 + |\psi_3|^2 + |\psi_4|^2 =
|G_0|^2$.\\
A grid size of $N_z
\times N_y = 1024^2$ elements is used and the initial wave
packet spread is set at $\sigma=48$, a fully relativistic particle
($m=0$) is considered.\\ In these simulations, we impose $g=0$ and $g=1000$ and
vary the initial energy of the wave packet $k\equiv k_z = k_y$ in
order to inspect the effect of this parameter on the wave packet
separation, which, in turn, informs on the effective mass acquired by the
up and down propagating modes.

\begin{figure}[htbp]
\centering
$g=0$ \hspace{2.5cm} $g=1000$\\ \vspace{0.2cm}
\includegraphics[scale=0.25]{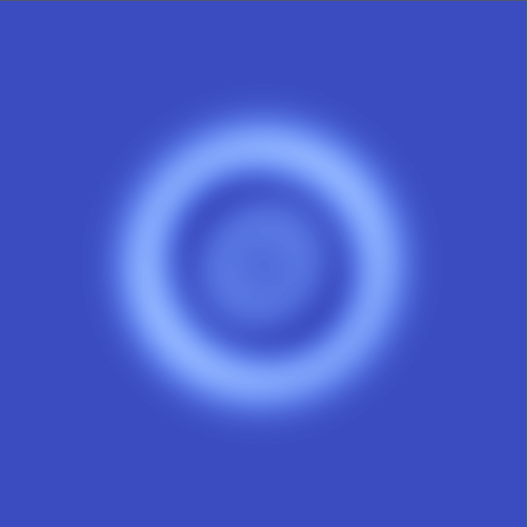}
\includegraphics[scale=0.25]{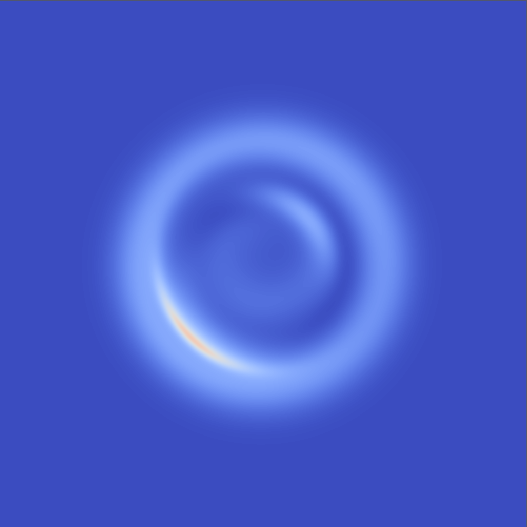}\\ \vspace{0.2cm}
\includegraphics[scale=0.25]{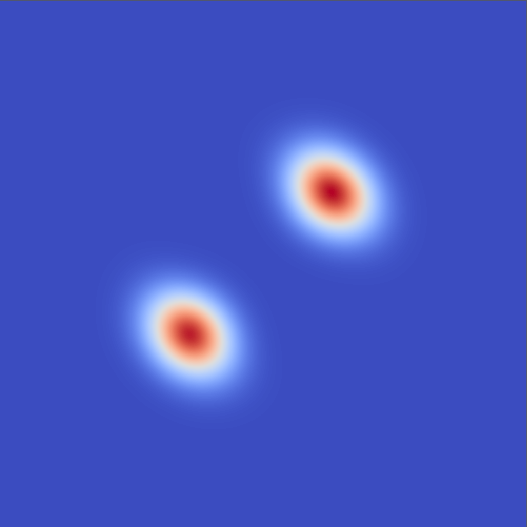}
\includegraphics[scale=0.25]{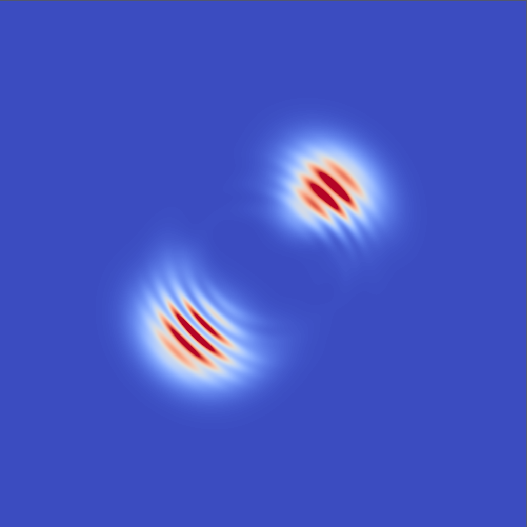}\\ \vspace{0.2cm}
\includegraphics[scale=0.25]{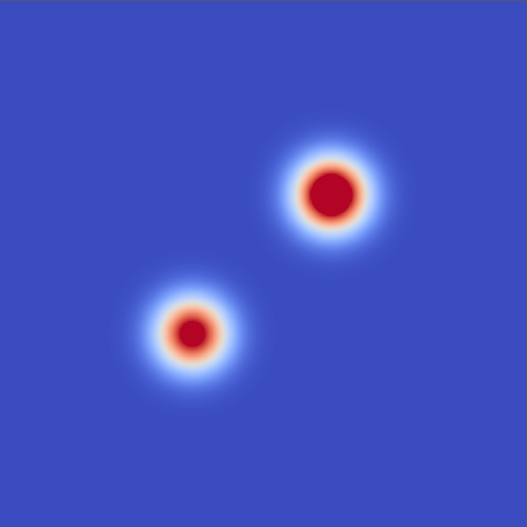}
\includegraphics[scale=0.25]{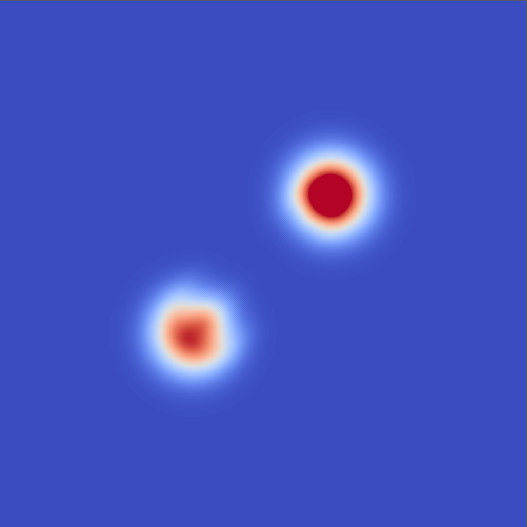}\\
\caption{Wave packet density $\rho = |\psi_1|^2 + |\psi_2|^2 + |\psi_3|^2 + |\psi_4|^2$ for the scheme solving Dirac equation with NJL
  interaction. These snapshots are taken at time $t=200$ for $k=0.004$, $0.04$ and $0.4$ (from top to bottom). The
figure shows how the initial energy affects the separation
phenomenon. In particular, at low energy, $k=0.004$,  no splitting of the wavepackest is observed.
}
\label{fig:varying_k}
\end{figure}

In \figref{varying_k}, the wave function density at time
$t=200$ for $k=0.004$, $0.04$ and $0.4$ is shown for $g=0$ and
$g=1000$, respectively.
The figure shows that sufficient energy, $k>0.004$, is needed to observe the splitting
of the wavepacket. The effects of non-linear interactions, fringes and distortions,
are also well visible in the right column of Figure 3,    
A quantitative analysis in the one-dimensional case led to satisfactory agreement with
asymptotic solutions for the dynamic mass as a function of the interaction strength $g$.
A similar analysis in two spatial dimensions remains to be developed.

\section{Summary and outlook}
\label{sec:summary}

Summarizing, we have reviewed discretizations of the Dirac equation and described their mapping into QW's. These relations may allow the solution of the Dirac equation on quantum computers. In the first part, a general argument is given, using the operator splitting method. Then, the QLB scheme is studied within the same perspective and a similar relation is found. We have also shown that a similar structure remains in curved space, using a scheme based on a finite volume formulation, with the important caveat
that the exact nature of the streaming operator, typical of QLB, is no longer preserved. Rather, one sees the appearance of the residency matrix, which characterizes the fraction of spinor which is left in the cell after one step in the time evolution. This scheme, along with its generalization to many dimensions, will be studied in future work.


\begin{backmatter}

\section*{Competing interests}
  The authors declare that they have no competing interests.

\section*{Author's contributions}
SS had the idea to link QLB to QW, and to develop a QLB-like method for the curved-space Dirac equation. He also participated in the calculation and development of the numerical methods. FFG introduced the operator splitting method and participated in the calculations and development of the numerical methods. 
SP developed the multi-dimensional extensions and performed the numerical calculations. 
All authors read and approved the final manuscript.  

\section*{Acknowledgements}
One of the authors (SS) is very grateful to F. Debbasch for introducing him to the
notion of quantum walks. He also wishes  to express gratitude to Marcelo Alejandro Forets for
organizing a very informative workshop on Relativistic Quantum Walks, where the ideas behind this paper
have been first drafted out.

\bibliographystyle{bmc-mathphys} 
\bibliography{bibliography}    

\end{backmatter}

%
%
%
%
%
%
%
%
%
%
%
%
%
%

\end{document}